\documentclass[11pt]{article}                     
\usepackage[osf,sc]{mathpazo} 
\def\be{\begin{equation}}
\def\ee{\end{equation}}
\def\bea{\begin{eqnarray*}}
\def\eea{\end{eqnarray*}}
 
\begin{document}                                                  
\vspace*{-2cm}
\begin{flushright} 
ANL-HEP-PR-12-31\\
\end{flushright}
\begin{center}
{\Large Ternary Plots for Neutrino Mixing Visualization}
 \vskip 0.5 cm
{Cosmas  K. Zachos}

\normalsize  {zachos@anl.gov}
\vskip 0.1  cm
 {\it High Energy Physics Division, Argonne National Laboratory\\
Argonne, IL 60439-4815, USA}
\vskip 1.1 true cm 
\noindent 
{\bf Abstract} 
{\small\begin{flushleft}  
The visualization advantages of ternary plots are illustrated for the PMNS neutrino mixing matrix. Unitarity constraints are incorporated automatically, in part, since barycentric plots of this type  allow 
three variables with a fixed sum to be plotted as mere points inside an equilateral triangle on a plane. 
\end{flushleft}    }
\end{center}    

\hrule 

\section{Introduction} 
The weak-charged-current mixing of quarks through the unitary CKM matrix and leptons through the essentially identical PMNS matrix are normally parametrized in terms of angles, which however, are often not
easy to visualize, consistently bound by data, and compare to theoretical conjectural schemes. Likewise,
the flavor content of neutrino mass eigenstates $\nu_{1,2,3}$ is often depicted by ``popsicle plots",
where the relative three flavor contents are depicted by different color sections adding up to a constant, 
but they do not make immediately evident how they contrast among themselves.

All three angles in quark mixing are {\em small},  $\theta_{12}\approx \pi/14 \approx  \lambda \approx 0.23$, $\theta_{23}\approx \lambda^2\approx 0.04$, $\theta_{13}\approx \lambda^3 \approx 0.01$  (in standard PDG conventions; from ref \cite{Zhang:2012bk}, for uniformity of treatment---no intense attention to accuracy is at issue here). By sharp contrast, for neutrino mixing, the reactor one is small, $\theta_{13}\approx 0.16 \pm 0.02$,  and {\em two are big}, the solar $\theta_{12}\approx (\pi/4 - 0.19) \pm 0.02$, and the atmospheric $\theta_{23}\approx (\pi/4 +0.00) \pm 0.09$ (estimates from ref \cite{Zhang:2012bk} for maximal CP violation).

In this short note, it is briefly pointed out that three variables constrained to sum to a constant, in this case 1,
are best plotted as positions inside an equilateral triangle on a plane, automatically incorporating this and other constraints geometrically, and possibly allowing for  {\em simultaneous} graphic visualization of many entities, here, many neutrinos, and  surveying their sensitivity to, e.g.  CP violation,  or comparing to models, without comparing specific angles of each.

Such plots have found applications in physics, of course, and are referred to as ``Dalitz plots" \cite{Dalitz:1954cq} for the three pion decays of kaons, and in neutrino physics by 
Fogli, Lisi, and Scioscia \cite{Fogli:1995uu}, as well as Petyt \cite{petyt}, without emphasis on their power to 
collectively account for and contrast larger numbers of states/points; this feature has been recognized in 
ref \cite{quigg} They are widespread as well in
physical chemistry, petrology, mineralogy, metallurgy, and other physical sciences; in population genetics, they are called  Gibbs triangles or de Finetti diagrams; in game theory, simplex plots.
\section{Ternary plots}
The PMNS unitary matrix $U$ mixes left-handed  fields of the three neutrino mass eigenstates $\nu_j$, $i=1,2,3$,  into lepton-flavor linear combinations $\nu_l$, $l=e,\mu,\tau$, named after the charged leptons they couple to,
\be
\nu_l=\sum_{j=1}^{3} U_{lj} ~\nu_j ~.
\ee
From unitarity, the normalized mass eigenstates can thus be expressed as composites of the flavor eigenstates,
\be
\nu_j=\sum_{l=e,\mu,\tau}  U^\dagger _{jl} ~ \nu_l ~.
\ee
 Since they are normalized, from unitarity, the 3-vectors of their absolute-squared coefficients, 
which represent their relative content of flavor, have the components of each sum to 1,
\be
N_j=(   | U_{ej} |^2, | U_{\mu j} |^2, | U_{\tau j} |^2).
\ee
In popsicle plots, these three components are depicted by different colors.

Also from unitarity, the first flavor components of these three vectors {\em also sum to one},  
$| U_{e1} |^2+ | U_{e2} |^2+ | U_{e3} |^2 =1$; and likewise for the second, $\mu$, and third, $\tau$.

Three-vectors whose components sum to one involve only two independent components, and so they span a
plane equilateral triangle. Consider the equilateral triangle with sides $a=2/\sqrt{3}$ and thus height 1.
Every point inside the equilateral triangle has its perpendiculars to the three sides sum to 1, often referred to as ``Viviani's theorem": this is self-evident, as the point divides the equilateral triangle into three triangles with the same base, $a$ (the sides of the equilateral), and the perpendiculars as 
their heights. Since the areas of the three triangles make up the area of the equilateral, the three heights sum up to  the height of the equilateral, for any point in the triangle.\\  
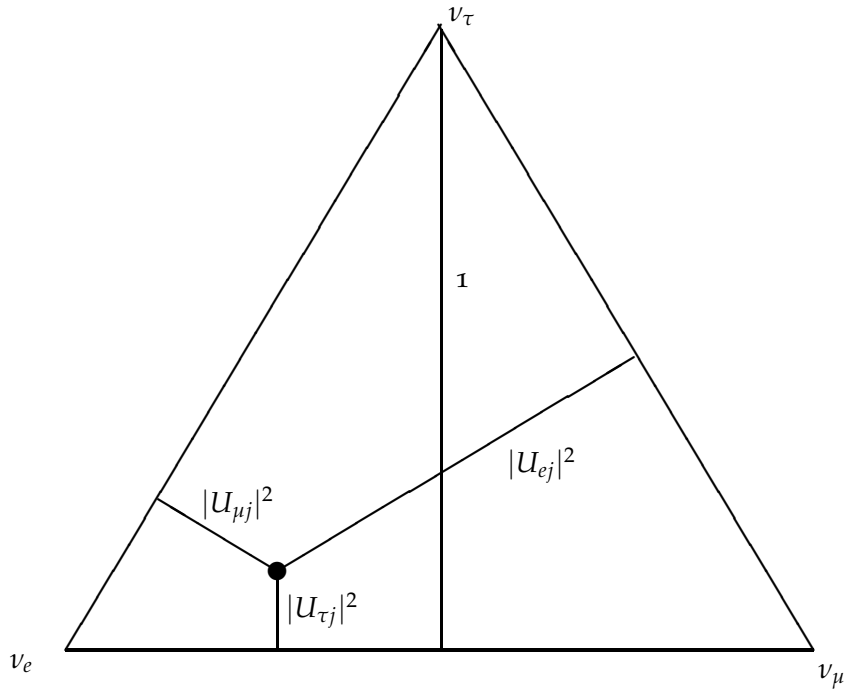
\begin{picture}(300,300)(100,0)  \thicklines
\put(120,40){\line(1,0){283}}
\put(120,40){\line(3,5){142}}
\put(403,40){\line(-3,5){142}}
\put(270,280){\makebox(0,0)[cc]{$\nu_\tau$}} 
\put(410,30){\makebox(0,0)[cc]{$\nu_\mu$}} 
\put(105,35){\makebox(0,0)[cc]{$\nu_e$ }} 
\put(200,70){\circle*{7}}
\put(200,70){\line(0,-1){30}}
\put(262,275){\line(0,-1){235}}
\put(200,70){\line(5,3){135}}
\put(200,70){\line(-5,3){45}}
\put(300,110){\makebox(0,0)[cc]{$|U_{ej}|^2$}} 
\put(185,95){\makebox(0,0)[cc]{$|U_{\mu j}|^2$}} 
\put(217,55){\makebox(0,0)[cc]{$|U_{\tau j}|^2$}} 
\put(270,180){\makebox(0,0)[cc]{1}} 
\end{picture}\\
Since the angles of the equilateral are all $\pi/3$, the Cartesian coordinates of the point representing the vector $N_j$  are $  (x,y)=\left ({2 |U_{\mu j}|^2+|U_{\tau j}|^2\over \sqrt{3}}  ,|U_{\tau j}|^2\right ) $. 
The origin $(x,y)=(0,0)$ then amounts to $\nu_e$.

\section {Neutrino and quark mixing} 
For a schematic illustration, we now plot the three neutrino mass eigenstate component vectors, with inputs  following the maximal CP violation case of  ref \cite{Zhang:2012bk}, 
\bea
N_1=(0.67,0.20,0.14),\\
N_2=(0.31,0.40,0.30),\\
N_3=(0.03,0.41,0.57)~,
\eea
with large solid circles; and contrast them to the (unrealistic) ``benchmark" bimaximal values \cite{Barger:1998ta} plotted  with large empty circles,
\bea
B_1=(1/2,1/4,1/4),\\
B_2=(1/2,1/4,1/4),\\
B_3=(~0~,~1/2,1/2)~,
\eea
and the erstwhile conjectural  tribamaximal \cite{Harrison:2002er} values with small empty circles,
\bea
T_1=(2/3,1/6,1/6),\\
T_2=(1/3,1/3,1/3),\\
T_3=(~0~,~1/2,1/2) ~.
\eea
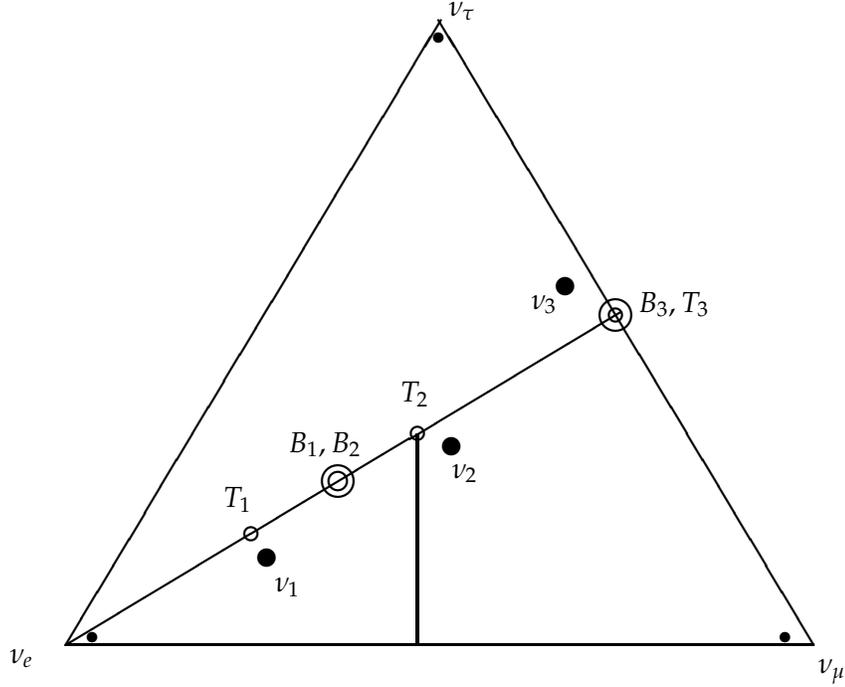
\begin{picture}(300,300)(100,0)  \thicklines
\put(120,40){\line(1,0){283}}
\put(120,40){\line(3,5){142}}
\put(403,40){\line(-3,5){142}}
\put(270,280){\makebox(0,0)[cc]{$\nu_\tau$}} 
\put(410,30){\makebox(0,0)[cc]{$\nu_\mu$}} 
\put(105,35){\makebox(0,0)[cc]{$\nu_e$ }} 
\put(120,40){\line(5,3){210}}
\put(309,176){\circle*{7}}
\put(301,169){\makebox(0,0)[cc]{$\nu_3$}} 
\put(350,169){\makebox(0,0)[cc]{$B_3,T_3$}} 
\put(328,165){\circle{5}}
\put(328,165){\circle{12}}
\put(253,120){\circle{5}}
\put(253,120){\line(0,-1){80}}
\put(252,135){\makebox(0,0)[cc]{$T_2$}} 
\put(266,115){\circle*{7}}
\put(271,105){\makebox(0,0)[cc]{$\nu_2$}} 
\put(223,102){\circle{7}}
\put(223,102){\circle{12}}
\put(218,116){\makebox(0,0)[cc]{$B_1,B_2$}} 
\put(190,82){\circle{5}}
\put(196,73){\circle*{7}}
\put(204,62){\makebox(0,0)[cc]{$\nu_1$}} 
\put(185,95){\makebox(0,0)[cc]{$T_1$}} 
\put(392,43){\circle*{4}}
\put(130,43){\circle*{4}}
\put(261,270){\circle*{4}}
\end{picture}\\
Both bi-and tribimaximal mixing paradigms require a vanishing $\theta_{13}$ and maximal $\theta_{23} = \pi/4$. Thus, they lie on the $\nu_\mu 
\leftrightarrow \nu_\tau$ approximate symmetry axis, also plotted to guide the eye. The middle tribamaximal point $T_2$ is the center of the equilateral triangle.

Also for contrast, the small solid circles near the vertices of the triangle 
represent the three quarks, where $\nu_e, \nu_\mu, \nu_\tau \mapsto d',s',b'$ is implied. Since the largest angle, the  Cabbibo angle, is of the order of magnitude of $\theta_{13}$,  they are clustered near the vertices of the triangle by amounts comparable to the offset of $\nu_3$ from the triangle side. Such collective contrasts would be unyieldy in tabular or popsicle plots.

Assuming unitarity, variation of   the hypothetical CP-violation phases $\delta_{CP}$  will generate 
small motions on short curves of the three points $N_j$ inside the triangle in unison: recall from the previous section that the unitarity constrains the sum of the verticals of all three points $N_j$ on each side of the triangle  separately  to also equal one. The plot may thus be of utility in surveying the collective sensitivity of such determinations on CP violation. Perhaps more significantly, the important task of comparing with scores of more realistic \cite{albr} theoretical texture schemes could be addressed more efficiently and compactly in a graphical manner, rather than a tabular form.
\section*{Acknowledgments}
Discussions with M Goodman, D Reyna, Z Djurcic, and Y Keung are gratefully acknowledged.
The submitted manuscript has been created by UChicago Argonne, LLC, Operator of Argonne National Laboratory (“Argonne”). Argonne, a U.S. Department of Energy Office of Science laboratory, is operated under Contract No. DE-AC02-06CH11357. The U.S. Government retains for itself, and others acting on its behalf, a paid-up nonexclusive, irrevocable worldwide license in said article to reproduce, prepare derivative works, distribute copies to the public, and perform publicly and display publicly, by or on behalf of the Government.

\end{document}